\documentstyle[12pt]{article}
\newskip\humongous \humongous=0pt plus 1000pt minus 1000pt

\newif\ifdtup

\def\beq{\begin{equation}}
\def\eeq{\end{equation}}

\def\beqn{\begin{eqnarray}}
\def\eeqn{\end{eqnarray}}
\relax

\def\G2{{\; \rm GeV/}c^2}
\def\G{\; \rm GeV}

\def\dotx{\dotx{\dot\overline{x}}}

\relax

\begin{document}
\begin{titlepage}
\begin{flushright}
       {\normalsize    OU-HET 250  \\ hep-th 9610***\\
             October,~1996  \\}
\end{flushright}
\vfill
\begin{center}
  {\large \bf  $N=2$  Supermultiplet of   Currents
  \\  and Anomalous Transformations 
      \\ in Supersymmetric Gauge Theory }\footnote{This work is
 supported in part by  Grant-in-Aid for  Scientific Research
(07640403)
from
 the Ministry of Education, Science and Culture, Japan.}

\vfill

         {\bf H.~Itoyama},~~{\bf M. Koike}  \\
            and \\

              {\bf H.~Takashino}
\vfill
        Department of Physics,\\
        Graduate School of Science, Osaka University,\\
        Toyonaka, Osaka, 560 Japan\\
      
\end{center}
\vfill
\begin{abstract}

  We examine some properties of supermultiplet 
 consisting of  the $U(1)_{J}$ current, 
  extended supercurrents, energy-momentum tensor
 and the central charge in $N=2$ supersymmetric Yang-Mills theory.
The superconformal improvement requires adding another supermultiplet
  beginning with  the $U(1)_{R}$ current.
 We determine  the anomalous (quantum mechanical) supersymmetry transformation
 associated with the central charge  and the energy-momentum tensor
 to one-loop order.

\end{abstract}
\vfill
\end{titlepage}

\section{Introduction}

  It is well-known that  the $U(1)_{R}$ current, the supercurrent
 $S_{\alpha}^{a}$ and the energy-momentum tensor $T^{ab}$ form a
 supermultiplet of currents in theories with $N=1$
 supersymmetry\cite{FZ}. (For a review, see\cite{W}  for example).
  In the pure super Yang-Mills case,  this supermultiplet is
\beq
J_{\alpha \dot{\alpha}}^{(V)} \equiv W_{\alpha} \bar{W}_{\dot{\alpha}} ,
\eeq
  whereas in  the scalar Wess-Zumino model it is
\beq
\label{eq:swz}
 J_{\alpha \dot{\alpha}}^{(S)} \equiv -\frac{1}{3}
 \bar{D}_{\dot{\alpha}} \Phi^{\dagger} D_{\alpha}\Phi
 -\frac{2}{3}
  \Phi^{\dagger}i \left( \stackrel{\leftrightarrow}{\partial}
\right)_{\alpha \dot{\alpha}}\Phi \;\;\;.
\eeq
 The second term in eq.~(\ref{eq:swz}) is necessary for the superconformal
 improvement $\sigma \cdot S^{a}=0,~T^{a}_{~a} =0$ .
 The corresponding situation in theories with extended supersymmetry,
 namely, the $N=2$ counterpart of supermultiplet of currents  does not
 seem to have been discussed  actively before except through
  coupling to $N=2$ supergravity \cite{SG}.
   In this note,  we wish to add a few things on this point.
 
We first note that  
 it is  the $U(1)_{J}$ current which is related to the energy momentum tensor
 and the density ( albeit the total
  derivative \cite{OW})  of the central charge  via the $N=2$ supersymmetry
 transformations.  This current is known to be anomaly free.
 The $U(1)_{R}$  current, which is anomalous,  comes in when
  one improves the supercurrent and  the energy-momentum tensor.
 This is presented in the next section.  In section three, 
we examine  anomalous (quantum mechanical) properties  of 
 the supersymmetry transformation \cite{IR} acting on the  $U(1)_{J}$ current.
    Operating the transformation twice, we  are able to derive a quantum
 mechanical property   assoicated with the  central charge of the
   extended supersymmetry algebra and that of the energy-momentum tensor. 
 The details of our computation as well as its implications will
  be discussed elsewhere.
  We follow the notation of \cite{WB}.

\section{Classical Transformation Law  of Currents}

 It is most convenient to begin  our discussion with the abelian case;
 we can then organize  the entire $N=2$ supermultiplet of currents in
 terms of the $N=2$ chiral and antichiral superfields;
\beqn
  {\cal A} &=& \Phi + i \sqrt{2} \eta W - \frac{1}{4} \eta \eta \bar{D}^{2}
 \Phi^{\dagger}  \nonumber \\
  {\cal A}^{\dagger} &=& \Phi^{\dagger} - i \sqrt{2} \bar{\eta}
 \bar{W} - \frac{1}{4} \bar{\eta} \bar{\eta} D^{2} \Phi \;\;\;.
\eeqn
Define 
\beqn
\label{eq:VJ}
 {\cal V}_{J}^{a}  \equiv - \frac{1}{2} (D_{\theta} {\cal A}) \sigma^{a}
  \bar{D}_{\theta} {\cal A}^{\dagger}  
   +  \frac{1}{2} (D_{\eta} {\cal A}) \sigma^{a}
  \bar{D}_{\eta} {\cal A}^{\dagger}   \;\;,
\eeqn  
   where $D_{\xi^{(i)}}$ is  the ordinary superderivative\cite{WB} 
  with respect ot the Grassmann number $\xi^{(i)}$
   with $\xi^{(1)}= \theta$, $\xi^{(2)}= \eta$.
  We find that   eq.~(\ref{eq:VJ})  contains $U(1)_{J}$ current
\beq
  J_{J}^{a} \equiv  - \psi \sigma^{a} \bar{\psi} + \lambda
 \sigma^{a} \bar{\lambda} \;\;\;.
\eeq
  as   the lowest componet.   This current is anomaly free.
  Higher components  are found  to be
\beqn
 &&-\sqrt{2}i \psi \sigma^{a} \bar{\sigma}^{b} \theta \partial_{b} A^{*}
  - v_{bc} \theta \sigma^{bc} \sigma^{a} \bar{\lambda} \;\;,\;\; \theta\;
 {\rm term}\;\;,    \label{eq:susy}  \\
&& +\sqrt{2}i \lambda \sigma^{a} \bar{\sigma}^{b} \eta \partial_{b} A^{*}
  - v_{bc} \eta \sigma^{bc} \sigma^{a} \bar{\psi} \;\;,\;\; \eta\;
 {\rm term}\;\;,  \label{eq:etasusy}    \\
&& -2 \sqrt{2} i \partial_{b}(A^{*} v^{ba}) -\sqrt{2}
 \epsilon^{abcd} \partial_{b} (A^{*} v_{cd}) \;\;, \;\; \eta \theta
\;  {\rm terms}
 \;\;,  \label{eq:center}  \\
&& -i \psi \sigma^{a} \partial^{b} \bar{\psi}+i\partial^{b} \psi \sigma^{a}
\bar{\psi}  -i \lambda\sigma^{a} \partial^{b} \bar{\lambda}
 +i \partial^{b} \lambda \sigma^{a} \bar{\lambda}  \nonumber \\
&&  -2 \partial^{a} A \partial^{b}A^{*} -2 \partial^{b} A \partial^{a} A^{*}
 -2 v^{b}_{c}v^{ac}  +2i \epsilon^{abcd} \partial_{c} A \partial_{d}A^{*}
 + \tilde{v}^{b}_{c}v^{ac}  \nonumber \\
&& - \tilde{v}^{a}_{c} v^{bc}
 + g^{ab} ( 2 \partial_{c} A \partial^{c} A^{*} +
 \frac{1}{2} v^{cd}v_{cd})
 \;, \; \theta \sigma_{a} \bar{\theta},\; \eta \sigma_{a} \bar{\eta}
\; {\rm terms} \;\;, \label{eq:emtensor}
\tilde{v}^{ab} \equiv \frac{i}{2}\epsilon^{abcd}{v}_{cd} \;\;\;,
 \nonumber\\
\eeqn
  which  are respectively the supercurrent $S^{a}_{(1)}$, 
 the current for extended supersymmetry $S^{a}_{(2)}$ , the density of
  the central  charge $C^{a}$  and the energy-momentum tensor $T^{ab}$.
  ( We have used equations of motion.  The remaining components
  are  either zero or a total derivative.)

 Eqs.~(\ref{eq:susy}), (\ref{eq:etasusy}) do not obey
 $\sigma \cdot S_{(i)} = 0\;,\; (i=1,2)$ condition.
  Neither does eq.~(\ref{eq:emtensor}) obey  the tracelessness condition.
 To make them obey  these conditions, ${\it i.e.}$ to  carry out
 the superconformal improvement,  we introduce  another $N=2$ 
 supermultiplet of currents:
\beqn 
 {\cal V}_{R}^{a} \equiv - \frac{1}{2} D_{\theta} {\cal A} \sigma^{a}
  \bar{D}_{\theta} {\cal A}^{\dagger}
   - \frac{1}{2} D_{\eta} {\cal A} \sigma^{a}\bar{D}_{\eta} {\cal A}^{\dagger}
    -2i {\cal A} \stackrel{\leftrightarrow}{\partial^{a}}
    {\cal A}^{\dagger} \;\;\;.
\eeqn
The lowest component  is  the $R$-current
\beqn
   J_{R}^{a} \equiv 2i (  A^{\dagger} \partial^{a} A
 - (\partial^{a} A^{\dagger})  A)
 - \psi \sigma^{a} \bar{\psi}- \lambda \sigma^{a} \bar{\lambda} \;\;\;.
\eeqn
 The higher components  do not seem to have straightforward interpretation.
  These can be used, however, for the superconformal improvement. Note that
  the $\theta$ and $\eta$ components  of $ {\cal V}_{J}^{a}$ and those of
 ${\cal V}_{R}^{a}$ satisfy 
\beqn
\left. {\cal V}_{J}^{a} \right|_{\theta = \bar{\xi} \bar{\sigma}^{a}}
 &=& - \left. {\cal V}_{J}^{a} \right|_{\eta = \bar{\xi} \bar{\sigma}^{a}}
 = 2 \sqrt{2} i \psi \sigma^{b} \bar{\xi} \partial_{b} A^{*}  \\
\left.  {\cal V}_{R}^{a} \right|_{\theta = \bar{\xi} \bar{\sigma}^{a}}
 &=&  \left. {\cal V}_{R}^{a} \right|_{\eta = \bar{\xi} \bar{\sigma}^{a}}
 = 4 \sqrt{2}i \psi \sigma \bar{\xi}\partial_{b} A^{*}\;\;\;.
\eeqn
  The trace of the $\theta \sigma^{b} \bar{\theta}$ term and
   that of the $\eta\sigma^{b}\bar{\eta}$ term
 in ${\cal V}_{J}^{a}$  are respectively $4\partial_{a}A \partial^{a} A^{*}$
 and  $-4\partial_{a}A \partial^{a} A^{*}$  whereas
  the  trace of the $\theta\sigma^{b}\bar{\theta}$ term and
  that of the $\eta\sigma^{b}\bar{\eta}$ term
  in ${\cal V}_{R}^{a}$  are $8\partial_{a}A \partial^{a} A^{*}$.
  Knowing these,  let us define
\beqn
\label{eq:imp}
 {\cal V}_{imp (1)}^{a} \equiv  {\cal V}_{J}^{a} -
 \frac{1}{2} {\cal V}_{R}^{a}  \;\;, \;\;
{\cal V}_{imp (2)}^{a} \equiv  {\cal V}_{R}^{a} + \frac{1}{2}
 {\cal V}_{R}^{a} \;\;\;.
\eeqn
 The first one ${\cal V}_{imp (1)}^{a}$ accomplishes  the superconformal
 improvement for ${\cal S}^{a}_{(1)}$ and $T_{ab}$. The second one  does
  for ${\cal S}^{a}_{(2)}$ and $T_{ab}$.
 
To conclude, it is $U(1)_{J}$  which is connected to
 the center of  the supersymmetry algebra via the supersymmetry
 transformations.  The superconformal improvement  is done by either one of the
 two in eq.~(\ref{eq:imp}).

  Having known the basic algebraic structure, let us turn to the nonabelian
 case.  No $N=2$ superfield formulation useful for our purpose
  seems to be known \cite{GSW,M,W}.  We carry out straightforward computation
 by components. 
  It is straightforward to derive $S_{(i)}^{a}$
 as an extended supertransformation on $J_{J}^{a}$.
\beqn 
 \delta_{(i)}J_{J}^{a} &=& \xi^{(i)}  {\cal S}_{(i)}^{a} +
 \bar{\xi}^{(i)} \bar{{\cal S}}_{(i)}^{a} \;\;, \nonumber \\
 {\cal S}_{(1) \alpha}^{a} &=& i tr \left\{ \sqrt{2}
 \left( \sigma^{b} \bar{\sigma}^{a}
 \psi \right)_{\alpha} D_{b} A^{*} -i \left(\sigma^{bc} \sigma^{a}
 \bar{\lambda}\right)_{\alpha} v_{bc}  \right.   \nonumber \\
  &&\;\;\;\;
 \left. -\frac{1}{2} \left( \sigma^{a} \bar{\lambda}\right)_{\alpha}
 \left[ A^{*}, A \right] \right\}   \;\;, \nonumber \\
\bar{ {\cal S}}_{(1)}^{a~\dot{\alpha}} &=& -i tr \left\{ \sqrt{2}
 \left( \bar{\sigma}^{b} \sigma^{a}
 \bar{\psi} \right)^{\dot{\alpha} } D_{b} A -i \left(\bar{\sigma}^{bc}
\bar{\sigma}^{a} \lambda \right)^{\dot{\alpha} } v_{bc} \right.  \nonumber \\
 &&\;\;\;\;
 \left. + \frac{1}{2} \left( \bar{\sigma}^{a} \lambda\right)^{\dot{\alpha}}
 \left[ A^{*}, A \right] \right\} \;\;,   \nonumber \\
 {\cal S}_{(2)\;\alpha}^{a} &=&
 -i tr \left\{ + \sqrt{2} \left( \sigma^{b} \bar{\sigma}^{a}
 \lambda \right)_{\alpha} D_{b} A^{*} +i \left(\sigma^{bc} \sigma^{a}
 \bar{\psi}\right)_{\alpha} v_{bc}  \right.   \nonumber \\
&& \;\;\;\; \left. +\frac{1}{2} \left( \sigma^{a} \bar{\psi}\right)_{\alpha}
 \left[ A^{*}, A \right] \right\}  \;\;,  \nonumber \\
\bar{ {\cal S}}_{(2)}^{a~\dot{\alpha}} &=& i tr \left\{ + \sqrt{2}
 \left( \bar{\sigma}^{b} \sigma^{a}
 \bar{\lambda} \right)^{\dot{\alpha} } D_{b} A + i \left(\bar{\sigma}^{bc}
\bar{\sigma}^{a} \psi \right)^{\dot{\alpha} } v_{bc}  \right.   \nonumber \\
&& \;\;\;\;
 \left. - \frac{1}{2} \left( \bar{\sigma}^{a} \psi \right)^{\dot{\alpha}}
 \left[ A^{*}, A \right] \right\}\;\;.
\eeqn
 Transforming once more, we obtain the energy-momentum tensor $T^{ab}$
  and the density $C^{a}$ of the central charge.
\beqn
 -2 \xi^{1}\sigma_{b} \bar{\xi}^{1} T^{ab} 
  \equiv \delta_{1}\bar {\delta_{2}} J_{J}^{a} =
 \xi^{1} \sigma_{b} \bar{\xi^{1}} (-2) \left[ tr \left\{ - D^{a}A^{*}D^{b}A 
 - D^{b}A^{*}D^{a}A 
   \right. \right. \nonumber \\
 \left.  \left.    -i \bar{\psi} \bar{\sigma}^{a}D^{b} \psi
-i \lambda
 \sigma^{a}D^{b} \bar{\lambda} -v^{ac}v_{bc}  \right\} -g^{ab} {\cal L} \right]
 + \xi \sigma_{b} \bar{\xi} tr \left\{ -2i \epsilon^{abcd} D_{c}A D_{d}A^{*}
 \right. \nonumber \\
 \left.  + \frac{1}{2} \epsilon^{abcd}v_{cd} \left[ A^{*}, A \right] \right\}
 +  \xi \sigma_{b} \bar{\xi} tr \left\{ v^{b}_{\;c}\tilde{v}^{ac} -
\tilde{v}^{b}_{\;c} v^{ac} \right\} \;\;\;. \nonumber\\
 \tilde{v}^{ab} \equiv \frac{i}{2}\epsilon^{abcd}{v}_{cd} \;\;\;.
 \nonumber\\
  C^{a} \equiv \delta_{1} \delta_{2}  J_{J}^{a} = 
  tr \left\{ 2 \sqrt{2} i D_{b} \left( A^{*} v^{ba} \right) +  \sqrt{2}
 \epsilon^{abcd} D_{d} \left(A^{*} v_{bc} \right) \right\} \;\;\;. 
\eeqn


\section{Quantum Mechanical Transformations} 

 In the remainder of this note,  we examine how some of these currents written
  in the interaction picture receive quantum corrections by studying
 supersymmetry transformations discussed above.  This will be done
 to one-loop.
  As  currents are in general composite operators and  the ones considered
 here are not gauged, it is perfectly consistent that the quantum mechanical
 counterpart of the classical transformations contains  new operators.
  For that reason, this term may be referred to as anomalous transformations. 
  The corresponding calculation in the $N=1$ case has been given in \cite{IR}.
  Let us write   generically 
\beqn
 \delta_{\theta} {\bf J}_{J}^{a}(x \mid \epsilon)   =
 \theta {\bf S}_{1}^{a}(x \mid \epsilon) + 
 \theta  {\bf \Delta}_{1}^{a}(x \mid \epsilon)  \nonumber \\
 \delta_{\eta} {\bf J}_{J}^{a}(x \mid \epsilon)   =
 \eta {\bf S}_{2}^{a}(x \mid \epsilon) + \eta 
 {\bf \Delta}_{2}^{a}(x \mid \epsilon) \nonumber \\
  \bar{\delta}_{\theta}  \delta_{\theta} {\bf J}_{J}^{a}(x \mid \epsilon)   =
 \bar{\theta} \theta {\bf T}^{a}(x \mid \epsilon) + 
  {\bf \Delta}_{EM}^{a}(x \mid \epsilon ;  \theta, \bar{\theta} ) \nonumber \\ 
 \delta_{\eta} \delta_{\theta} {\bf J}_{J}^{a}(x \mid \epsilon)   =
 \eta \theta {\bf C }^{a}(x \mid \epsilon) +
 {\bf \Delta}_{center}^{a}(x \mid \epsilon ; \eta, \theta ) \;\;.
\eeqn
   Here the $U(1)_{J}$ current in the Heisenberg picture
  regularized by gauge invariant point splitting is
\beqn
 {\bf J}_{J}^{a}(x) = tr \left[ \bar{ {\bf \Psi}} ( x+\frac{\epsilon}{2})
 \bar{\sigma}^{a} U_{\epsilon} {\bf \Psi} (x- \frac{\epsilon}{2})U_{-\epsilon}
 - \bar{{\bf \Lambda}} ( x+\frac{\epsilon}{2})
 \bar{\sigma}^{a} U_{\epsilon} {\bf \Lambda}
 (x- \frac{\epsilon}{2})U_{-\epsilon}  \right] \nonumber \\
 U_{\epsilon} \equiv  P \exp \left[ -i\frac{1}{2}
 \int_{x - \frac{\epsilon}{2}}^{x + \frac{\epsilon}{2}} dx^{a}
 v_{a}(x)\right] \;\;\;.
\eeqn
  We will use bold face letters for the operators in the Heisenberg picture.

It is easy to see  that ${\bf \Delta}_{i}^{a}(x \mid \epsilon) = 0
 \;\;i =1,2$  by antisymmetry under ${\bf \Psi} \leftrightarrow {\bf \Lambda}$.
  We will now compute  ${\bf \Delta}_{center}^{a}(x \mid \epsilon ; 
 \eta, \theta)$  and 
 ${\bf \Delta}_{EM}^{a}(x \mid \epsilon; \theta, \bar{\theta})$ to
one-loop order. 

  We find
\beqn
\label{eq:acenter}
 {\bf \Delta}_{center}^{a}(x \mid \epsilon ; 
 \eta, \theta) &=& - \frac{i}{2} \epsilon ^b
  tr\left\{ 
-\sqrt{2} D_c{\bf A}^* \left(x+\frac{\epsilon}{2} \right) \left[ \eta
 \sigma _b  \bar{ {\bf \Psi}}\left( x\right) ,\theta \sigma ^c \bar{\sigma}^a
 {\bf \Psi} \left( x-\frac{\epsilon}{2}\right) \right] \right. \nonumber \\
& &- i\bar{{\bf \Psi}}\left( x+\frac{\epsilon}{2}\right)
 \bar{\sigma}^a \sigma ^{cd} \eta \left[ \theta \sigma _b \bar{ {\bf \Lambda}}
 \left( x\right) ,{\bf v}_{cd} \left( x -\frac{\epsilon}{2}\right)
  \nonumber  \right]\\
& &+\bar{{\bf \Psi} } \left( x+\frac{\epsilon}{2}\right)
 \bar{\sigma}^a \eta \left[ \theta \sigma_b \bar{{\bf \Lambda} }
 \left( x\right) ,
{\bf D}\left( x-\frac{\epsilon}{2}\right) \right] \nonumber  \\
& &-\sqrt{2} \theta \sigma_b \bar{\sigma}^c \eta \,
 \bar{{\bf \Psi}}
 \left( x+\frac{\epsilon}{2}\right) \left[D_c {\bf A}^* \left( x\right) ,
 \bar{\sigma}^a {\bf \Psi} \left( x-\frac{\epsilon}{2}
 \right)\right]  \nonumber \\
& &-\sqrt{2} D_c {\bf A}^* \left(x+\frac{\epsilon}{2} \right)
 \left[ \theta \sigma _b \bar{{\bf \Lambda} }\left( x\right) ,\eta \sigma ^c
 \bar{\sigma}^a {\bf \Lambda} \left( x-\frac{\epsilon}{2}
\right) \right]  \nonumber \\
& &+ i\bar{{\bf \Lambda}}\left( x+\frac{\epsilon}{2}\right)
 \bar{\sigma}^a \sigma ^{cd} \theta \left[ \eta \sigma _b \bar{{\bf \Psi}}
 \left( x\right) ,{\bf v}_{cd} \left( x -\frac{\epsilon}{2}\right)
   \right]  \nonumber \\
& &-\bar{{\bf \Lambda}} \left( x+\frac{\epsilon}{2}\right)
 \bar{\sigma}^a \theta \left[ \eta \sigma_b \bar{{\bf \Psi}} \left( x\right) ,
 {\bf D}\left( x-\frac{\epsilon}{2}
\right) \right] \nonumber  \\
& &\left. +\sqrt{2} \theta \sigma_b \bar{\sigma}^c \eta \,
 \bar{{\bf \Lambda}} \left( x+\frac{\epsilon}{2}\right)
 \left[D_c {\bf A}^* \left( x\right) , \bar{\sigma}^a {\bf \Lambda}
 \left( x-\frac{\epsilon}{2} \right)\right]\right\} \;.
\eeqn
Let
\beqn
  \lim_{\epsilon \rightarrow 0}
  {\bf \Delta}_{center}^{a}(x \mid \epsilon ; 
 \eta, \theta) =  \sum_{a=1}^{4} I_{a}
  + \sum_{a=1}^{4} I_{a}^{\prime} \;\;,
\eeqn 
 where $I_{a}$ $a=1\sim 4$ and $I_{a-4}^{\prime}$ $a=5\sim 9$  represent
  the  respective contributions from  the $a$-th line in eq.~(\ref{eq:acenter})
 evaluated to one-loop in the interaction picture.

 Space does not permit us to present  one-loop computation of all $I_{a}$'s
 and $I_{a}^{\prime}$'s we have carried out.  We just demonstrate our
calculation in a sample term:
\beqn 
 I_{1} &=& \lim_{\epsilon \rightarrow 0}- \frac{i}{2} (-\sqrt{2}) 
 tr \left\{ T^{r} [ \eta \sigma_{b}
 \bar{\psi} (x), T^{s} ] \right\} \epsilon^{b} \nonumber \\
  &\times&  \int d^{4}y \frac{1}{g^{2}} \frac{1}{\sqrt{2}} tr \left\{ T^{t}
[ T^{u}, T^{v}] \right\} < \partial_{c}A^{(r)*}(x+\frac{\epsilon}{2}) A^{(t)}
(y)>  \nonumber \\
 &\times&  \theta \sigma^{c} \bar{\sigma}^{a}
 < \psi^{(s)}(x- \frac{\epsilon}{2}) 
 \bar{\psi}^{(u)} (y) > \bar{\lambda}^{(v)}(y)   \nonumber \\
&=&\lim_{\epsilon \rightarrow 0} \frac{1}{2} g^{2} 
 tr_{adj}\left( \eta\sigma_{b} \bar{\psi}(x)T_{(adj)}^{v}
 \right) \int d^{4}y \int \frac{d^{4}p }{(2 \pi)^{4}i}
 \frac{d^{4} q}{(2\pi)^{4}i} \theta \sigma^{c} \bar{\sigma}^{a} \sigma^{d}
 \bar{\lambda}^{(v)}(y)  \nonumber \\
&\times& \epsilon^{b} \frac{p_{c}q_{d}}{p^{2}q^{2}} e^{i(p-q)(x-y)}
 e^{i(p+q)\frac{\epsilon}{2}} \nonumber \\
&=&  i\left( \frac{1}{2} \right)^{2} \frac{g^{2}}{32 \pi^{2}}
 tr_{(adj)} \eta\sigma_{b} \bar{\psi} (x)
 \left( \theta \sigma^{b} \bar{\sigma}^{a} \sigma^{c} \partial_{c}
 \bar{\lambda}(x) -  \theta \sigma^{c} \bar{\sigma}^{a} \sigma^{b}
 \partial_{c} \bar{\lambda}(x) \right) \;\;.
\eeqn
  We have used
\beqn
 \lim_{\epsilon \rightarrow 0}\int \frac{d^{4}p }{(2 \pi)^{4} i}
 \frac{d^{4} q}{(2\pi)^{4} i}  \epsilon^{b} \frac{p_{c}q_{d}}{p^{2}q^{2}}
 e^{i(p-q)(x-y)} e^{i(p+q)\frac{\epsilon}{2}} \nonumber \\
=  \frac{i}{64 \pi^{2}} \left( \delta^{b}_{c} \frac{\partial}{\partial x^{d}}
 - \delta^{b}_{d} \frac{\partial}{\partial x^{c}} \right)
 \delta^{(4)}(x-y) \;. 
\eeqn

  Collecting all contributions, we find the final answer: 
\beqn
  \lim_{\epsilon \rightarrow 0} {\bf \Delta}_{center}^{a}(x \mid \epsilon) =
   i (\frac{1}{2})^{2} \frac{g^{2}}{32\pi^{2}} \left[ \partial_{b} \left(
 - \eta \theta \bar{\psi} \bar{\sigma}^{ab} \bar{\lambda} -
 2 \theta \sigma^{ab} \eta
   \bar{\psi} \bar{\lambda} \right)   \right. \nonumber \\
\left. + \frac{3}{2} \eta \theta ( \bar{\psi} \partial^{a}
 \bar{\lambda} - \bar{\lambda}
 \partial^{a} \bar{\psi} )
        - \eta \sigma^{ab} \theta
 ( \bar{\psi} \bar{\sigma}_{b} \sigma_{c} \partial^{c} \bar{\lambda} +
\bar{\lambda} \bar{\sigma}_{b} \sigma_{c} \partial^{c} \bar{\psi} )
 \right] \;.
\eeqn 

  As for the energy-momentum tensor,  a similar calculation yields
\beqn
  {\bf \Delta}_{EM}^{a}(x \mid \epsilon ;  \theta, \bar{\theta} )= \theta
 \sigma_{b}\bar{\theta}  {\bf \Delta}_{EM}^{ab}(x \mid \epsilon ) \nonumber \\ 
  \lim_{\epsilon \rightarrow 0}   {\bf \Delta}_{EM}^{ab}(x \mid \epsilon )
 = -i \left( \frac{1}{2}\right)^{2} \frac{g^{2}}{64 \pi^{2}}
  tr \left[ 3 \eta^{ab} \left( (\partial^{c} \lambda) \sigma_{c} \bar{\lambda}
 - \lambda \sigma_{c} \partial^{c} \bar{\lambda} \right) \right. \nonumber \\
 \left. + \left( (\partial^{a} \lambda) \sigma^{b} \bar{\lambda}
- \lambda\sigma^{b} \partial^{a} \bar{\lambda} \right)
-  \left( (\partial^{b} \lambda) \sigma^{a} \bar{\lambda}
- \lambda\sigma^{a} \partial^{b} \bar{\lambda} \right)
 + i \epsilon^{abcd} \partial_{c} ( \lambda \sigma_{d} \bar{\lambda}) \right]
 \;\;.
\eeqn


\section {Acknowledgements}
  We thank  Alyosha Morozov  for discussions on a related topic.

\end{document}